\def\year{2022}\relax
\documentclass[letterpaper]{article} 
\usepackage{aaai22}  
\usepackage{times}  
\usepackage{helvet}  
\usepackage{courier}  
\usepackage[hyphens]{url}  
\usepackage{graphicx} 
\usepackage{natbib}  
\usepackage{caption} 
\DeclareCaptionStyle{ruled}{labelfont=normalfont,labelsep=colon,strut=off} 
\usepackage{algorithm}
\usepackage{algorithmic}

\usepackage{bm}
\usepackage{colortbl}
\usepackage{arydshln}
\usepackage{multirow}
\usepackage{multicol}
\usepackage{xcolor}
\usepackage{mathtools}
\usepackage{amsmath}
\usepackage{amsfonts}

\usepackage{newfloat}
\usepackage{listings}
\floatstyle{ruled}
\newfloat{listing}{tb}{lst}{}
\floatname{listing}{Listing}
\title{Deep Reinforcement Learning for Cryptocurrency Trading: Practical Approach to Address Backtest Overfitting}

\author{
    Berend Jelmer Dirk Gort\textsuperscript{\rm 1}\equalcontrib\thanks{B. Gort finished this project as a research assistant at Columbia University.},
    Xiao-Yang Liu\textsuperscript{\rm 1}\equalcontrib,
    Jiechao Gao\textsuperscript{\rm 2}, \\
    Shuaiyu Chen \textsuperscript{\rm 3},
    Christina Dan Wang \textsuperscript{\rm 4}
}
\affiliations{
    \textsuperscript{\rm 1}Department of Electrical Engineering, Columbia University\\
    \textsuperscript{\rm 2}Department of Computer Science, University of Virginia\\
    \textsuperscript{\rm 3}Department of Finance at Krannert School of Management, Purdue University\\
    \textsuperscript{\rm 4}Department of Business and Finance, New York University (Shanghai)\\
}

\begin{document}

\maketitle

\begin{abstract}

Designing profitable and reliable trading strategies is challenging in the highly volatile cryptocurrency market. Existing works applied deep reinforcement learning methods and optimistically reported increased profits in backtesting, which may suffer from the \textit{false positive} issue due to model overfitting. In this paper, we propose a practical approach to address backtest overfitting for cryptocurrency trading using deep reinforcement learning. First, we formulate the detection of backtest overfitting as a hypothesis test. Then, we train DRL agents, estimate the probability of overfitting, and reject overfitted agents, increasing the chance of good trading performance. Finally, on 10 cryptocurrencies over a testing period from 05/01/2022 to 06/27/2022 (during which the crypto market \textbf{crashed twice}), we show that less overfitted deep reinforcement learning agents have a higher return than that of more overfitted agents, an equal weight strategy, and the S\&P DBM Index (market benchmark), offering confidence in possible deployment in a real market.
\end{abstract}


\section{Introduction}
\label{section: Introduction}

A profitable and reliable trading strategy in the cryptocurrency market is critical for hedge funds and investment banks. Deep reinforcement learning methods prove to be a promising approach \cite{Fang2022}, including crypto portfolio allocation \cite{Jiang2018, ang2022asset} and trade execution \cite{hambly2021recent}.  However, three major challenges prohibit the adoption in a real market: 1) the cryptocurrency market is highly volatile \cite{yang2019price}; 2) the historical market data have a low signal-to-noise ratio \cite{Conrad2018}; and 3) there are large fluctuations (e.g., market crash) in the cryptocurrency market. 

Existing works may suffer from the backtest overfitting issue, which is a \textit{false positive} issue. Many methods \cite{xiong2018practical, liu2021finrl, Yang2020} adopted a walk-forward method and optimistically reported increased profits in backtesting. The walk-forward method divides data in a training-validation-testing split, but using a single validation set can easily result in model overfitting. Another approach \cite{Jiang2018} considered a $K$-fold cross-validation method (leaving one period out) with an underlying assumption that the training and validation sets are drawn from an IID process, which does not hold in financial tasks. For example, \cite{liu2021risks} showed that there is a strong momentum effect in crypto returns. Finally, DRL algorithms are highly sensitive to hyperparameters \cite{li2021finrl}, resulting in high variability of DRL algorithms' performance \cite{Clary2019, Henderson2018, Mania2018}. A researcher may get `lucky' during the backtest process and obtain an \textit{overoptimistic} agent (overfitted one). A key question from practitioners is that ``\textit{does the trained agent generalize to different market situations?}"

This paper proposes a practical approach to address the backtest overfitting issue. Researchers in the cryptocurrency trading niche submit papers containing overfitted backtest results. A practical and quantitative approach for detecting model overfitting is valuable. First, we formulate the detection of backtest overfitting as a hypothesis test. Such a test employs an estimated probability of overfitting to determine whether a trained agent is acceptable. Second, we provide detailed steps to estimate the probability of backtest overfitting. If the probability exceeds a preset threshold, we reject it. Finally, on 10 cryptocurrencies over a testing period from 05/01/2022 to 06/27/2022 (during which the crypto market \textbf{crashed twice}), we show that less overfitted deep reinforcement learning agents have a higher return than that of more overfitted agents, an equal weight strategy, and the S\&P DBM Index (market benchmark), offering confidence in possible deployment in a real market.  We hope DRL practitioners may apply the proposed method to ensure that their chosen agents are not just false positive results. 



The remainder of this paper is organized as follows. We first review related works. Then, we describe the cryptocurrency trading task, and further propose a practical approach to address the backtest overfitting issue. Finally, we present performance evaluations that are followed by conclusions.


\section{Related Works}
\label{section: Related Works}

Existing works can be classified into three categories: 1) backtest using the walk-forward method, 2) backtest using the cross-validation method, and 3) backtest with hyperparameter tuning. 

\subsection{Backtest Using Walk-Forward}

The Walk-Forward (WF) method is the most widely applied backtest scheme. WF trains a DRL agent over a training period and then evaluates its performance over a subsequent validation period. However, WF validates in one market situation, which can easily result in overfitting \cite{Agarwal2021, Lin2021, Chan2019}. WF may not reflect future performance, since the validation period can be biased, e.g., a significant market uptrend. Therefore, we want to train and validate under various market situations to avoid overfitting and make the agent more robust.

\subsection{Backtest Using Cross-Validation}

Conventional methods \cite{Jiang2018} used $K$-fold cross-validation (KCV) to backtest agents' trading performance. The KCV method partitions the dataset into $K$ subsets, generating $K$ folds. Then, for each trial, select one subset as the testing set and the rest $K-1$ subsets as the training set. However, there are still risks of overfitting. First, a $K$-fold cross-validation method splits data by drawing from an IID process, which is a bold assumption for financial markets \cite{robinson1994time}. Second, the testing set generated by a cross-validation method could have a substantial bias \cite{LopezdePrado2018}. Finally, informational leakage is possible because the training and testing sets are correlated \cite{farokhi2020modelling}. The researcher should be cautious about leaking testing knowledge into the training set.

We employ an improved approach. Existing research overlooked the backtest overfitting issue and gave the false impression that DRL-based trading strategies may be ready to deploy on markets \cite{Varoquaux2021, Bouthillier2021, Dodge2020}. WF only tests a single market situation with high statistical uncertainty, and KCV takes a false IID assumption without control for leakage. Therefore, we employ a combinatorial cross-validation method that tracks the \textit{degree of overfitting} during the backtest process. This method simulates a large number of market situations and estiamtes the probability of overfitting.


\subsection{Backtest with Hyperparameter Tuning}

DRL agents are highly sensitive to hyperparameters, as can be seen from the implementations of Stable Baselines3 \cite{raffin2021stable}, RLlib \cite{liang2018rllib}\cite{liaw2018tune}, UnityML \cite{Juliani2018} and TensorForce \cite{tensorforce}. The selection of hyperparameters takes a lot of time and strongly influences the learning result. Several cloud platforms provide hyperparameter tuning services.

In finance, FinRL-Podracer \cite{li2021finrl} employed an evolutionary strategy for training a trading agent on a cloud platform, which ranks agents with different hyperparameters and then selects the one with the highest return as the resulting agent. Essentially, this method carried out a hyperparameter tuning process. Further, ElegantRL-Podracer \cite{liu2021elegantrl} generalized it to a tournament-based method that is a highly scalable and cloud-native scheduling scheme for a GPU cloud.


\section{Cryptocurrency Trading Using Deep Reinforcement Learning}
\label{section: Cryptocurrency Trading Using Deep Reinforcement Learning}

First, we model a cryptocurrency trading task as a Markov Decision Process (MDP). Then, we build a market environment using historical market data and describe the general setting for training a trading agent. Finally, we discuss the backtest overfitting issue.

\begin{figure}
    \centering
    \includegraphics[width=0.8\linewidth]{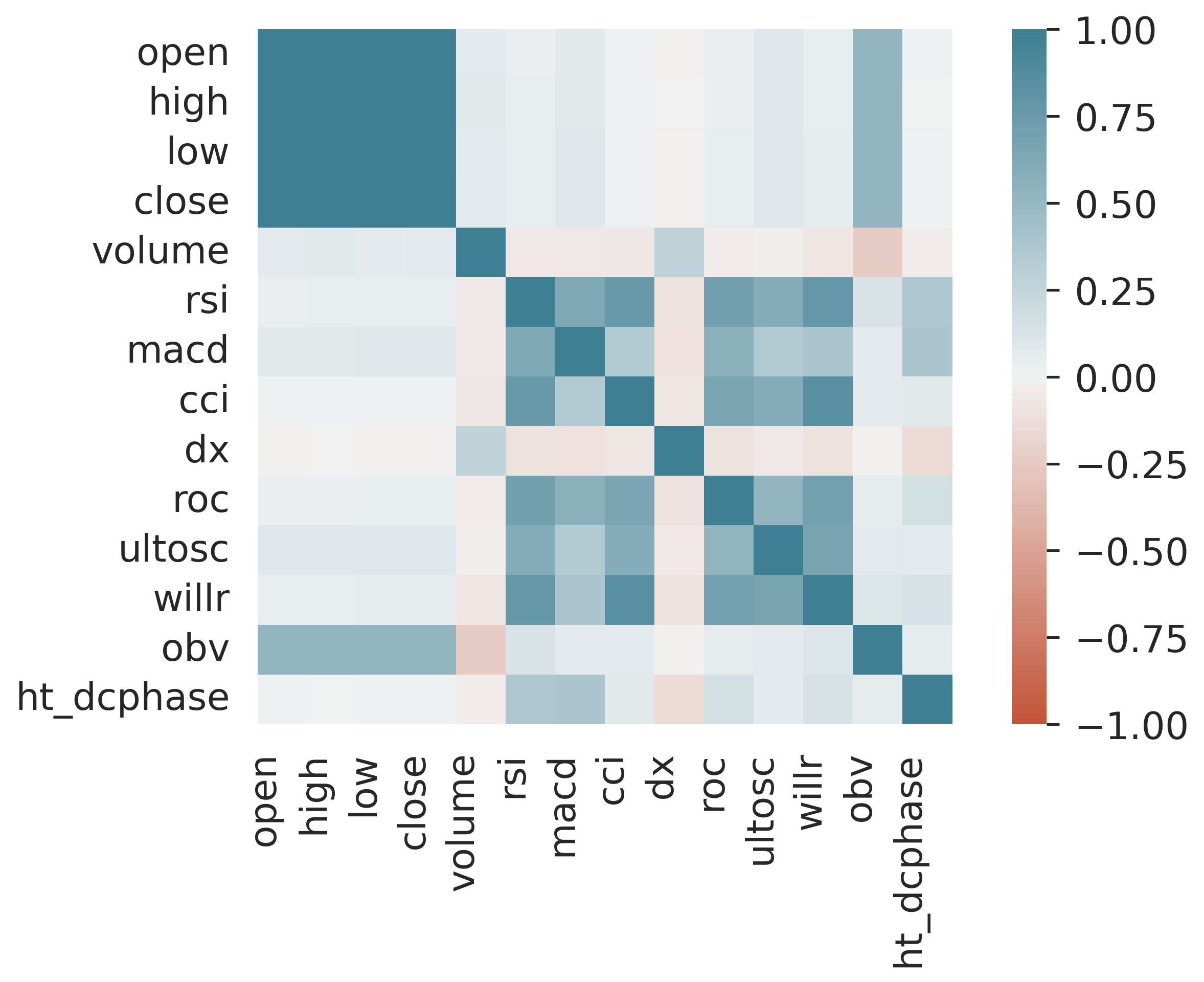}
    \caption{The correlation matrix of features.}
    \label{fig:high_correlation_filter}
\end{figure}

\subsection{Modeling Cryptocurrency Trading}
\label{subsection: MDP Model for Crypto Trading}

 Assuming that there are $D$ cryptocurrencies and $T$ time slots, $t=0,1,...,T-1$. We use a deep reinforcement learning agent to make trading actions, which can be either \emph{buy, sell, or hold}. An agent observes the market situation, e.g., prices and technical indicators, and takes actions to maximize the cumulative return. 
We model a trading task as a Markov Decision Process (MDP) as follows
\begin{itemize}
    \item State $\bm{s}_t = [b_t, \bm{h}_t, \bm{p}_t, \bm{f}_t] \in \mathbb{R}^{1+ (I+2)D}$, where $b_t \in \mathbb{R}_{+}$ is the cash amount in the account,  $\bm{h}_t \in \mathbb{R}_{+}^{D}$ denotes the share holdings, $\bm{p_t}  \in \mathbb{R}_{+}^{D}$ denotes the prices at time $t$, $\bm{f}_t \in \mathbb{R}^{I D}$ is a feature vector for $D$ cryptocurrencies and each has $I$ technical indicators, and $\mathbb{R}_{+}$ denotes non-negative real numbers.
    \item Action $\bm{a}_t \in  \mathbb{R}^D$ changes the non-negative share holdings $\bm{h}_t \in \mathbb{R}_{+}^{D}$, i.e., $\bm{h}_{t+1} = \bm{h}_{t} + \bm{a}_t$, where positive actions increase $\bm{h}_{t}$,  negative actions decrease $\bm{h}_{t}$, and zero actions keep ${h}_{t}$ unchanged.  
    \item Reward $r(\bm{s_{t}, \bm{a_{t}}, \bm{s_{t+1}}}) \in \mathbb{R}$ is defined as a return when taking action $\bm{a}_t$ at state $\bm{s}_t$ and arriving at a new state $\bm{s}_{t+1}$. Here, we set it as the change of portfolio value, i.e., $r(\bm{s_{t}, \bm{a_{t}}, \bm{s_{t+1}}}) = v_{t+1} - v_t$, where $v_t = \bm{p}_t^{\top} \bm{h}_{t} + b_{t} \in \mathbb{R}_+$.
    \item Policy $\pi(\bm{a}_t | \bm{s}_t )$ is the trading strategy, which is a probability distribution over actions at state $\bm{s}_t$.
\end{itemize}

We consider $15$ features that are used by existing papers \cite{xiong2018practical, zhang2020deep, Yang2020,liu2021finrl}, e.g., open-high-low-close-volume (OHLCV) and $9$ technical indicators. Over the training period (from 02/02/2022 to 04/30/2022, as shown in Fig. \ref{fig:Data_Split}), we compute Pearson correlations of the features and obtain a correlation matrix in Fig. \ref{fig:high_correlation_filter}. We list the $9$ technical indicators as follows:

\begin{itemize}
\label{list: state-space}
    \item Relative Strength Index (RSI) measures price fluctuation. 
    \item Moving Average Convergence Divergence (MACD) is a momentum indicator for moving averages.
    \item Commodity Channel Index (CCI) compares the current price to the average price over a period.
    \item Directional Index (DX) measures the trend strength by quantifying the amount of price movement.
    \item The rate of change (ROC) is the speed at which variable changes over a period \cite{Gerritsen2020}.
    \item Ultimate Oscillator (ULTSOC)  measures the price momentum of an asset across multiple timeframes \cite{UgurGudelek2018}.
    \item Williams \%R (WILLR) measures overbought and oversold levels \cite{Ni2009}.
    \item On Balance Volume (OBV) measures buying and selling pressure as a cumulative indicator that adds volume on up-days and subtracts volume on down-days \cite{Gerritsen2020}.
    \item The Hilbert Transform Dominant (HT) is used to generate in-phase and quadrature components of a detrended real-valued signal to analyze variations of the instantaneous phase and amplitude \cite{Nava2016}.
\end{itemize}

As shown in Fig. \ref{fig:high_correlation_filter}, if two features have a correlation coefficient exceeding $\pm 0.6$, we drop either one of the two. Finally, $I=6$ features are kept in the feature vector $\bm{f}_t \in \mathbb{R}^{6D}$, which are trading volume, RSI, DX, ULTSOC, OBV, and HT. Since the OHLC prices are highly correlated, $\bm{p}_t$ in $\bm{s}_t$ is chosen to be the close price over the period $[t, t+1]$. Note that the close price of period $[t, t+1]$ equals to the open price of period $[t+1, t+2]$. The feature vector $\bm{f}_t$ characterizes the market situation. For the case $D=10$, $\bm{s}_t$ has size $81$.  

\subsection{Building Market Environment}
\label{sec: market environment}

We build a market environment by replaying historical data \cite{Liu2022}, following the style of OpenAI Gym \cite{Brockman2016}. A trading agent interacts with the market environment in multiple episodes, where an episode replays the market data (time series) in a time-driven manner from $t=0$ to $t=T-1$. At the beginning ($t=0$), the environment sends an initial state $\bm{s}_0$ to the agent that returns an action $\bm{a}_0$. Then, the environment executes the action $\bm{a}_t$ and sends a reward value $r_t$ and a new state $\bm{s}_{t+1}$ to the agent, for $t=0,...,T-1$. Finally, $\bm{s}_{T-1}$ is set to be the terminal state.

The market environment has the following three functions:
\begin{itemize}
    \item \texttt{reset} function resets the environment to $\bm{s}_0 =[ b_0, \bm{h}_0, \bm{p}_0, \bm{f}_0]$ where $b_0$ is the investment capital and $\bm{h}_0=0$ (zero vector) since there are no shareholdings yet. 
    \item \texttt{step} function takes an action $a_t$ and updates  state $\bm{s}_t$ to $\bm{s}_{t+1}$. For $\bm{s}_{t+1}$ at time $t+1$, $\bm{p}_{t+1}$ and $\bm{f}_{t+1}$ are accessible by looking up the time series of market data, and update $b_{t+1}$ and $\bm{h}_{t+1}$ as follows:
    \begin{equation}
\label{eq: postive portfolio balance constraint}
\begin{split}
    b_{t+1} &=b_{t}- \bm{p}_{t}^{\top} \bm{a}_{t},\\
    \bm{h}_{t+1} &=\bm{h}_{t} + \bm{a}_{t}.
\end{split}
\end{equation}
    \item \texttt{reward} function computes $r(\bm{s_{t}, \bm{a_{t}}, \bm{s_{t+1}}}) = v_{t+1} - v_t$ as follows:
    \begin{equation}
\label{eq: reward function portfolio value}
r\left(\bm{s}_{t}, \bm{a}_{t}, \bm{s}_{t+1}\right)=\left(b_{t+1}+\bm{p}_{\bm{t}+\mathbf{1}}^{\top} \bm{h}_{\bm{t}+\mathbf{1}}\right)-\left(b_{t}+\bm{p}_{\bm{t}}^{\top} \bm{h}_{\bm{t}}\right).
\end{equation}
\end{itemize}

\textbf{Trading constraints}

1). \textbf{Transaction fees}. Each trade has transaction fees, and different brokers charge varying commissions. For cryptocurrency trading, we assume that the transaction cost is $0.3\%$ of the value of each trade.
Therefore, (\ref{eq: reward function portfolio value}) becomes
\begin{equation}
\label{eq: reward function portfolio value_transaction_fee}
r\left(\bm{s}_{t}, \bm{a}_{t}, \bm{s}_{t+1}\right)=\left(b_{t+1}+\bm{p}_{\bm{t}+\mathbf{1}}^{\top} \bm{h}_{\bm{t}+\mathbf{1}}\right)-\left(b_{t}+\bm{p}_{\bm{t}}^{\top} \bm{h}_{\bm{t}}\right) - c_t,
\end{equation}
and the transactions fee $c_t$ is
\begin{equation}
c_{t}=\bm{p}_t^{\top} |\bm{a}_{t}| \times 0.3 \%,
\end{equation}
where $|\bm{a}_{t}|$ means taking entry-wise absolute value of $\bm{a}_{t}$.

2). \textbf{Non-negative balance}. We do not allow short, thus we make sure that $b_{t+1} \in \mathbb{R}_+$ is non-negative,
\begin{equation}
\label{eq: postive portfolio balance constraint2}
    b_{t+1}=b_{t}+ \bm{p}_{t}^{\top} \bm{a}_{t}^{S} + \bm{p}_{t}^{\top} \bm{a}_{t}^{B} \geq 0,~~~\text{for}~~t = 0,..., T-1,
\end{equation}
where $\bm{a}_{t}^{S} \in \mathbb{R}_{-}^D$ and $\bm{a}_{t}^{B} \in  \mathbb{R}_{+}^D$ denote the selling orders and buying orders, respectively, such that $\bm{a}_t = \bm{a}_{t}^{S}+\bm{a}_{t}^{B}$. Therefore, action $\bm{a}_t$ is executed as follows: first execute the selling orders $\bm{a}_{t}^{S} \in \mathbb{R}_{-}^D$ and then the buying orders $\bm{a}_{t}^{B} \in  \mathbb{R}_{+}^D$; and if there is not enough cash, a buying order will not be executed. 

3). \textbf{Risk control}. The cryptocurrency market regularly drops in terms of market capitalization, sometimes even $\geq 70\%$. To control the risk for these market situations, we employ the Cryptocurrency Volatility Index, CVIX \cite{Bonaparte2021}. Extreme market situations increase the value of CVIX. Once the CVIX exceeds a certain threshold, we stop buying and then sell all our cryptocurrency holdings. We resume trading once the CVIX returns under the threshold. 

\subsection{Training a Trading Agent}

A trading agent learns a policy $\pi(\bm{a}_t | \bm{s}_t)$ that maximizes the discounted cumulative return $R=\sum_{t=0}^{\infty} \gamma^{t} r\left(\bm{s}_{t}, \bm{a}_{t}, \bm{s}_{t+1}\right)$, where $\gamma \in (0,1]$ is a discount factor and $r\left(\bm{s}_t, \bm{a}_t, \bm{s}_{t+1}\right)$ is given in (\ref{eq: reward function portfolio value_transaction_fee}). The Bellman equation gives the optimality condition for an MDP problem, which takes a recursive form as follows:
\begin{equation}
Q^{\pi}(\bm{s_t}, \bm{a_{t}})=r(\bm{s_{t}, \bm{a_{t}}, \bm{s_{t+1}}}) +\gamma \mathbb{E}_{\bm{a_{t+1}}}\left[Q^{\pi}(\bm{s_{t+1}, a_{t+1}})\right].
\end{equation}
There are dozens of DRL algorithms that can be adapted to crypto trading. Popular ones are TD3 \cite{Fujimoto2018}, SAC \cite{haarnoja2018soft}, and PPO \cite{schulman2017proximal}. 

Next, we describe a general flow of agent trading. At the beginning of training, we set hyperparameters such as the learning rate, batch size, etc. DRL algorithms are highly sensitive to hyperparameters, meaning that an agent's trading performance may vary significantly. We have multiple sets of hyperparameters in the training stage for different trials. Each trial trains with one set of hyperparameters and obtains a trained agent. Then, we pick the DRL agent with the best-performing hyperparameters and re-train the agent on the whole training data.

\subsection{Backtest Overfitting Issue}
\label{appendix: Updated Model Selection Algorithm}

\textbf{Backtest} \cite{LopezdePrado2018} uses historical data to simulate the market and evaluates the performance of an agent, namely, how would an agent have performed should it have been run over a past time period. Researchers often perform backtests by splitting the data into two chronological sets: one training set and one validation set. However, a DRL agent usually overfits an individual validation set that represents one market situation, thus, the actual trading performance is in question.

\textbf{Backtest overfitting} occurs when a DRL agent fits the historical training data to a harmful extent. The DRL agent adapts to random fluctuations in the training data, learning these fluctuations as concepts. However, these concepts do not exist, damaging the performance of the DRL agent on unseen states.

\begin{figure}
    \centering
    \includegraphics[width=0.9\linewidth]{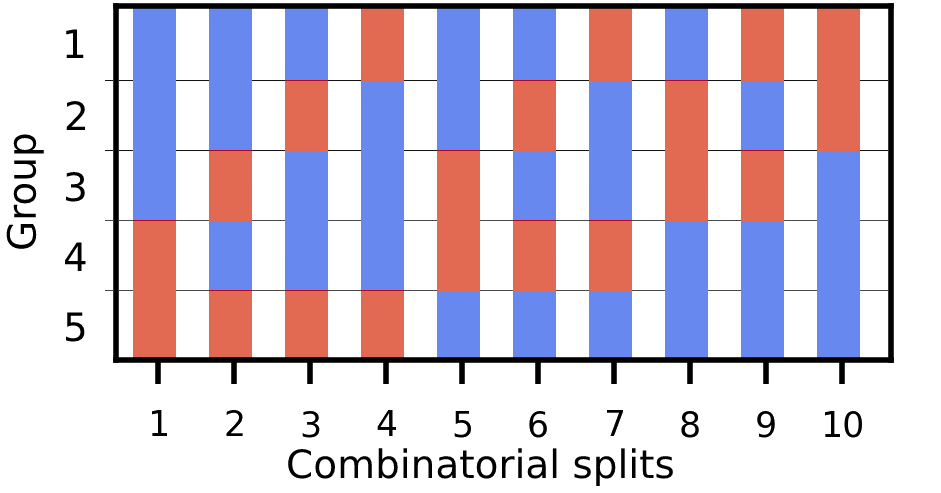}
    \caption{Illustration of combinatorial splits. Split the data into $N=5$ groups, $k=2$ groups for the train set (red) and the rest $N - k = 3$ groups for the second set (blue).}
    \label{fig:CV_splits_group.png}
\end{figure}


\section{Practical Approach to Address Backtest Overfitting}
\label{section: Statistical Verification of a Deep Reinforcement Learning Trading Strategy}

We propose a practical approach to address the backtest overfitting issue \cite{Bailey2017}. First, we formulate the problem as a hypothesis test and reject agents that do not pass the test. Then, we describe the detailed steps to estimate the probability of overfitting, $p \in [0,1]$.

\subsection{Hypothesis Test to Reject Overfitted Agents}


We formulate a hypothesis test to reject overfitted agents. Mathematically, it is expressed as follows
\begin{equation}\label{hypothesis_test}
\begin{cases}
    \mathcal{H}_0:&  p < \alpha, \quad \quad \text{NOT overfitted}, \\
    \mathcal{H}_1:&  p \geq \alpha, \quad \quad \text{overfitted}.
\end{cases}    
\end{equation}
where $\alpha > 0$ is the level of significance.

The hypothesis test (\ref{hypothesis_test}) is expected to reject two types of \textit{false-positive} DRL agents. 1) Existing methods may have reported overoptimistic results since many authors were tempted to go back and forth between training and testing periods. This type of information leakage is a common reason for backtest overfitting. 2). Agent training is likely to overfit since DRL algorithms are highly sensitive to hyperparameters \cite{Clary2019, Henderson2018, Mania2018}. For example, one can train agents with  PPO, TD3, or SAC algorithm and then reject agents that do not pass the test. We set the level of significance $\alpha$ according to the Neyman-Pearson framework  \cite{neyman1933testing}.

\subsection{Estimating the Probability of Overfitting}
\label{sec: Estimating Probability of Overfitting}

\textbf{Combinatorial Cross-validation} allows for training-and-validating on different market situations. Given a training time period, we perform the following steps:
\begin{itemize}
    \item \textbf{Step 1} (Training-validation data splits): as shown in Fig. \ref{fig:CV_splits_group.png}, divide the training period with $T$ data points into $N$ groups of equal size, $k$ out of $N$ groups construct a validation set and the rest $N - k$ groups as a training set, resulting in $J=\left(\begin{array}{c}N\\ N-k\end{array}\right)$ combinatorial splits. The training and validation sets have $(N-k)(T/N)$ and $T'=k(T/N)$ data points, respectively.
    \item \textbf{Step 2} (One trial of hyperparameters): set a new set of parameters for hyperparameter tuning.
    \item \textbf{Step 3}:  In each training-validation data split, we train an agent using the training set and then evaluate the agent's performance metric for each validation set, $i = 1, ..., H$. After training on all splits, we take the mean performance metric over all validation sets.
\end{itemize}

Step 2) and Step 3) constitute \textit{one hyperparameter trial}. Loop for $H$ trials and select the set of hyperparameters (or DRL agent) that performs the best in terms of mean performance metric over all splits. This procedure considers various market situations, resulting in the best-performing DRL agent over different market situations. However, a training process involving multiple trials will result in overfitting \cite{Bailey2014}.

We estimate the probability of overfitting using the return vector. The return vector is defined as $\bm{R}_t = v_{t} - v_{t-1}$, where $v_t$ is the portfolio value at time step $t$. We estimate the probability of overfitting via three general steps.

\begin{itemize}
    \item \textbf{Step 1}: For each hyperparameter trial, average the returns on the validation sets (of length $T'$) and obtain $\bm{R}_{\text{avg}} \in \mathbb{R}^{T'}$.
    \item \textbf{Step 2}: For $H$ trials, stack $\bm{R}_{\text{avg}}$ into a matrix $\bm{M} \in \mathbb{R}^{T'\times H}$. 
    \item \textbf{Step 3}: Based on $\bm{M}$, we compute the probability of overfitting $p$.
\end{itemize}

Consider a probability space $(\mathcal{T}, \mathcal{F}, \mathbb{P})$, where $\mathcal{T}$ represents the sample space, $\mathcal{F}$ the event space and $\mathbb{P}$ the probability space. A sample $c \in \mathcal{T}$ is a \textbf{split of matrix $\bm{M}$ across rows}. For instance, we split $\bm{M}$ into four subsets:
\begin{align}
\bm{M^1}, \bm{M^2}, \bm{M^3}, \bm{M^4} \in \mathbb{R}^{T'/4 \times H}.
\end{align}
A sample $c$ could be any split of $\bm{M}$, such as: 
\begin{equation}
    c = \begin{bmatrix}\bm{M^1}  \\ \bm{M^2}  \end{bmatrix}, ~ \text{IS set} \quad \quad\quad\quad\quad \bar{c} = \begin{bmatrix}\bm{M^3}  \\ \bm{M^4}  \end{bmatrix}, ~ \text{OOS set}
\end{equation}

\textit{An agent is overfitted if its best performance in the IS set has an expected ranking that is lower than the median ranking in the OOS set} \cite{Bailey2017}. For a sample $c \in \mathcal{T}$, let $\bm{R}^c \in \mathbb{R}^{H}$ and $\bm{\overline{R}^c} \in \mathbb{R}^{H}$ denote the IS and OOS performance of the columns of a sample $c$, respectively. We rank $\bm{R}^c$ and $\bm{\bar{R}}^c$ from low values to high values, resulting in $\bm{r}^c$ and $\bm{\bar{r}}^c$ (for example, [1, 3, 2], where 3 corresponds to the best performance metric). Define  $\epsilon$ as the index of the best-performing IS strategy. Then, we check the corresponding OOS rank $\bm{\bar{r}}^c[\epsilon$] and define a relative rank $\omega^c$ as follows:
\begin{equation}\label{def:relative_rank}
    \omega^c = \frac{\bm{\bar{r}}^c[\epsilon]}{H + 1}, ~\text{for}~c \in \mathcal{T}.
\end{equation}

Define a logit function as follows:
\begin{equation}\label{def:logit_function}
    \lambda^{c}=\ln \frac{\omega^c}{1-\omega^c},~\text{for}~c \in \mathcal{T}.
\end{equation}
If $\omega^c < 0.5$, we have $\lambda^c < 0$, meaning that the best strategy IS has an expected ranking lower than the OOS set, which is overfitting.  High logit values indicate coherence between IS and OOS performance, indicating a low level of overfitting. Finally, the probability of overfitting is computed as follows:
\begin{equation}\label{def:pbo}
    p = \int_{-\infty}^{0} f(\lambda) d \lambda,
\end{equation}
where $f(\lambda)$ denotes the distribution function of $\lambda$.

Finally, we discuss how to set the significance level $\alpha$. The relative OOS rank is a uniform distribution. Such a uniform's logit is normally distributed. In accordance with the Neyman-Pearson framework, we can set the level of significance $\alpha$. Because  $\alpha$ is a probability, it ranges between $0$ and $1$. Thus, if $\alpha=10\%$, we allow a 10\% probability of incorrectly rejecting the null hypothesis in favor of the alternative when the null hypothesis is true. 


\section{Performance Evaluations}
\label{section: Performance Evaluations}

First, we describe the experimental settings and the compared methods and metrics. Then, we show that the proposed method can help reject two types of overfitted agents. Finally, we present the backtest performance to verify that our hypothesis test helps increase the chance of good trading performance.

\subsection{Experimental Settings}
\label{section: Backtesting Settings}

We select $10$ cryptocurrencies with high trading volumes: AAVE, AVAX, BTC, NEAR, LINK, ETH, LTC, MATIC, UNI, and SOL. We assume a trade can be executed at the market price and ignore the slippage issue because a higher trading volume indicates higher market liquidity. 

\begin{figure}
    \centering
    \includegraphics[width=1\linewidth]{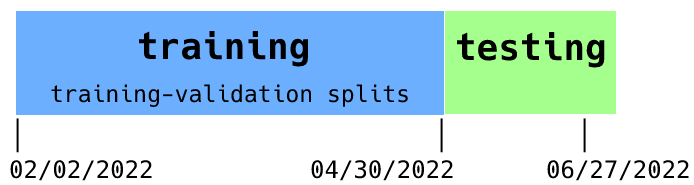}
    \caption{Data split: train and validate an agent in the blue period, and test its performance in the green period.}
    \label{fig:Data_Split}
\end{figure}

\textbf{Data split}: We use five-minute-level data from 02/02/2022 to 06/27/2022. We split it into a training period (from 02/02/2022 to 04/30/2022) and a testing period (from 05/01/2022 to 06/27/2022, during which the crypto market \textbf{crashed two times}), as shown in Fig. \ref{fig:Data_Split}. The training period splits further into IS-OOS sets for estimating $p$ as in the previous section.

\textbf{Training with combinatorial cross-validation}: there are in total $T = 25055 = 87 ~ (\text{days}) \times 24 ~(\text{hours}) \times 60/5 ~(\text{minutes}) - 1$ datapoints in the training time period, and  $T' = 16704 = 58 ~ (\text{days}) \times 24 ~(\text{hours}) \times 60/5 ~(\text{minutes})$ in the testing time period. We divide the training data set into $N= 5$ equal-sized subsets, each subset has 5011 data points. We perform the combinatorial cross-validation method with $N = 5$ and  $k=2$, where each data split has $k=2$ validation sets and $N-k = 3$ training sets. Therefore, the total number of training-validation splits is $J = 10$. 

\textbf{Hyperparameters}: we list six tunable hyperparameters in Table \ref{tab: hyperparameters}.  Their values are based on the implementations of Stable Baselines3 \cite{raffin2021stable}, RLlib \cite{liang2018rllib}, Ray Tune \cite{liaw2018tune}, UnityML \cite{Juliani2018} and TensorForce \cite{tensorforce}. There are in total $2700 = 5 \times 4 \times 5 \times 3 \times 3 \times 3$ combinations.

\textbf{Trials $H$}: for any distribution over a sample space with a finite maximum, the maximum of 50 random observations lies within the top 5\% of the actual maximum, with 90\% probability. Specifically, $1-(1-0.05)^H \geq 0.9$ requires $H \geq 50$ assuming the optimal region of hyperparameters occupies at least 5\% of the grid space.

\textbf{Volatility index CVIX} (Crypto VIX) \cite{Bonaparte2021}: the total market capitalization of the crypto market \textbf{crashed two times} in our testing period, namely, 05/06/2022 - 05/12/2022 and  06/09/2022 - 06/15/2022. We take the average CVIX value over those time frames as our threshold, $\text{CVIX}_t = 90.1$.

\textbf{Level of significance $\alpha$}: we allow a 10\% probability of incorrectly rejecting the null hypothesis in favour of the alternative when the null hypothesis is true (type I error), $\alpha=10\%$.

\begin{table*}
\centering
\caption{The hyperparameters and their values.}
\label{tab: hyperparameters}
\begin{tabular}{|l|l|l|}
\hline
\textbf{Hyperparameter} & \textbf{Description} & \textbf{Range} \\ \hline
\textbf{Learning rate} & Step size during the training process & $[3e^{-2}, 2.3e^{-2}, 1.5e^{-2}, 7.5e^{-3}, 5e^{-6}]$ \\ \hline
\textbf{Batch size} & Number of training samples in one iteration & $[512, 1280, 2048, 3080]$ \\ \hline
\textbf{Gamma $\gamma$} & Discount factor & $[0.95, 0.96, 0.97, 0.98, 0.99]$ \\ \hline
\textbf{Net dimension} & Width of hidden layers of the actor network & $[2^9 ,2^{10}, 2^{11}]$ \\ \hline
\textbf{Target step} & Explored target step number in the environment & $[2.5e^3, 3.75e^3, 5e^3]$  \\ \hline
\textbf{Break step} & Total timesteps performed during training & $[3e^4, 4.5e^4, 6e^4]$ \\ \hline
\end{tabular}%
\end{table*}

\subsection{Compared Methods and Performance Metrics}

We consider two cases where the proposed method helps reject overfitted agents.

\subsubsection{Conventional Deep Reinforcement Learning Agents}
\label{sec_conventional_agent}

First, the walk-forward method \cite{Park2020, Moody2001, Deng2017, Yang2020, Li2019} applies a training-validation-testing data split. On the same training-validation set, we train with $H=50$ different sets of hyperparameters, all with PPO algorithm \cite{schulman2017proximal} and then calculate $p$ \cite{Bailey2014}. Second, we train another conventional agent using the PPO algorithm and the $K$-fold cross-validation (KCV) method with $k=5$.

\subsubsection{Deep Reinforcement Algorithms with Different Hyperparameters}
\label{sec_diff_DRL_algos}

DRL algorithms are highly sensitive to hyperparameters, resulting in high variability of trading performance \cite{Clary2019, Henderson2018, Mania2018}. We use the probability of overfitting to measure the likelihood that an agent is overfitted. We tune the hyperparameters in Table \ref{tab: hyperparameters} for three agents, TD3 \cite{Fujimoto2018}, SAC \cite{haarnoja2018soft} and PPO, and calculate $p$ for each agent with each set of hyperparameters for $H=50$ trials.

\subsubsection{Performance Metrics}

We use three metrics to measure an agent's performance:
\begin{itemize}
  \item \textbf{Cumulative return} $R = \frac{v - v_0}{v_0}$, where $v$ is the final portfolio value, and $v_0$ is the original capital.
  \item \textbf{Volatility} $\text{V} = \text{std}(R_t)$, where $R_t = \frac{v_t - v_{t-1}}{v_{t-1}}$, and $t=0,...,T-1$.
  \item \textbf{Probability of overfitting $p$}. We split $\bm{M}$ into $14$ submatrices, analyze all possible combinations, and set the threshold $\alpha = 10\%$ \cite{neyman1933testing}.
\end{itemize}
The cumulative return measures the profits. The widely used volatility measures the degree of variation of a trading price series over time; the amount of risk.

\subsubsection{Benchmarks}

We compare two benchmark methods: an equal-weight portfolio and the S\&P Cryptocurrency Broad Digital Market Index (S\&P BDM Index). 

\begin{itemize}
    \item Equal-weight strategy: at time $t_0$, distribute the available cash $b_0$ equally over all $D$ available cryptocurrencies.
    \item S\&P BDM index \cite{spy_digital_indices2022}: the S\&P broad digital market index S\&P tracks the performance of cryptocurrencies with a market capitalization greater than \$10 million.
\end{itemize}

\subsection{Rejecting Conventional Agents}
\label{reject_conventional}

\begin{figure}
    \centering
    \includegraphics[width=1\linewidth]{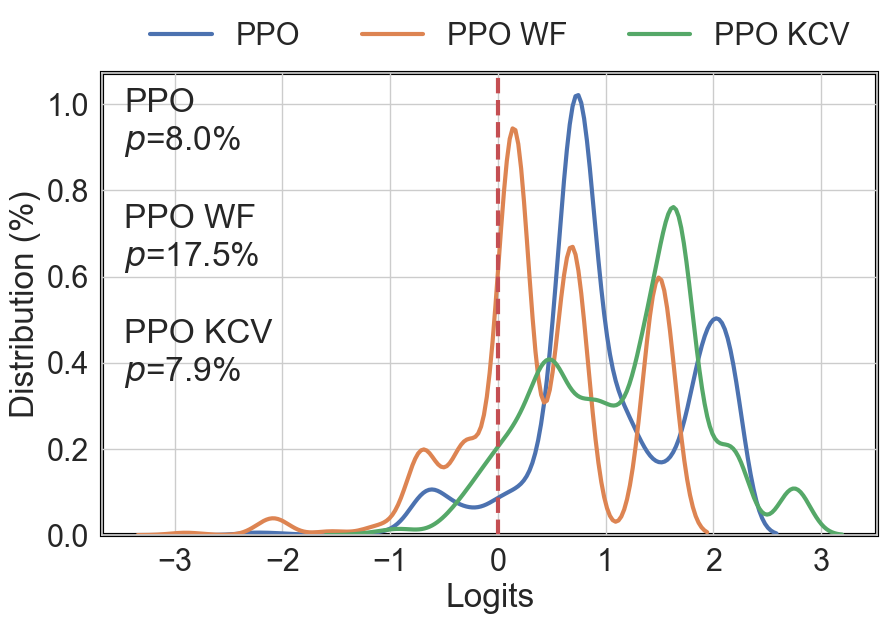}
    \caption{Logit distribution $f(\lambda)$ of three conventional agents.}
    \label{fig:multiple_logits_dist_53}
\end{figure}

Fig. \ref{fig:multiple_logits_dist_53} shows the logit distribution function $f(\lambda)$ for conventional agents described in Section \ref{sec_conventional_agent}. The area under $f(\lambda)$ for the domain $[-\infty, 0]$ is the probability of overfitting $p$. The peaks represent the outperforming DRL agents, as an agent's success is sensitive to the hyperparameter set. These outperforming trials deliver the same relative rank more often, and the logits are a function of the relative rank. We compare our PPO approach to conventional PPO WF and PPO KCV. The WF and KCV methods have $p_{\text{WF}} = 17.5\% > \alpha$ and $p_{\text{KCV}} = 7.9 \% < \alpha$, respectively. For the WF method, we accept the alternative hypothesis $\mathcal{H}_1$ and conclude that it is overfitting.


\begin{figure}
    \centering
    \includegraphics[width=1\linewidth]{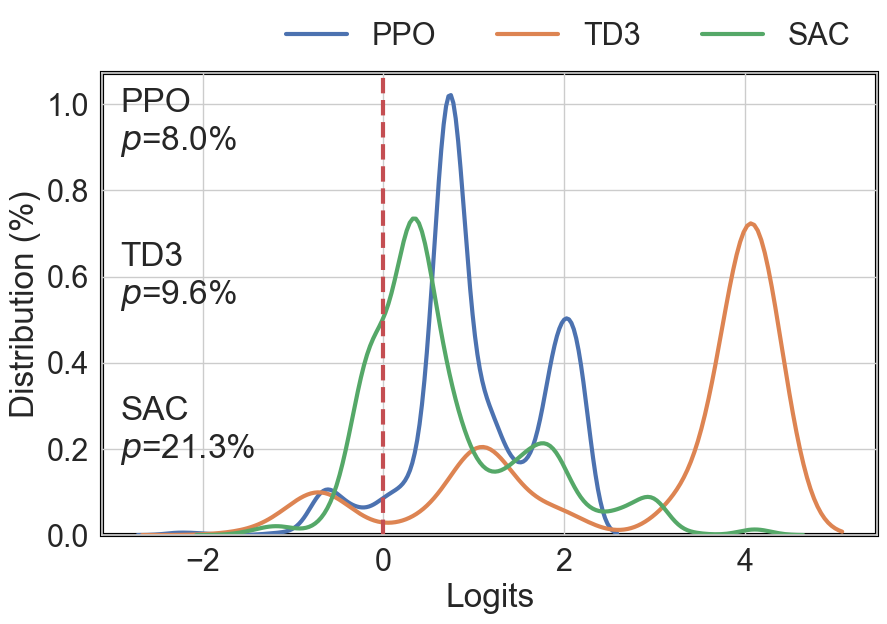}
    \caption{Logit distribution $f(\lambda)$ of three DRL agents.}
    \label{fig:multiple_logits_dist_54}
\end{figure}

\subsection{Rejecting Overfitted Agents}
\label{reject_drl_agents}


Table \ref{tab: hyperparameters_per_agent} presents the hyperparameters selected for each agent. Fig. \ref{fig:multiple_logits_dist_54} shows the logit density function for the DRL agents in Section \ref{sec_diff_DRL_algos}. The probabilities of overfitting are: $p_{\text{PPO}} = 8.0\% < \alpha$, $p_{\text{TD3}} = 9.6\% < \alpha$ and $p_{\text{SAC}} = 21.3\% > \alpha$. We accept alternative hypothesis $\mathcal{H}_1$ and conclude that SAC is overfitted. Finally, TD3 has a dominant part of the logit distribution at a high logit domain ($\approx 4$). High logit values indicate coherence between IS and OOS performance.

\begin{table}
\centering
\caption{Selected hyperparameters for each DRL algorithm.}
\label{tab: hyperparameters_per_agent}
\begin{tabular}{|l|l|l|l|}
\hline
\textbf{Hyperparameters} & \textbf{PPO} & \textbf{TD3} & \textbf{SAC} \\ \hline
\textbf{Learning rate} & 7.5e-3 & 3e-2 & 1.5e-2 \\ \hline
\textbf{Batch size} & 512 & 3080 & 3080 \\ \hline
\textbf{Gamma} & 0.95 & 0.95 & 0.97 \\ \hline
\textbf{Net dimension} & 1024 & 2048 & 1024 \\ \hline
\textbf{Target step} & 5e4 & 2.5e3 & 2.5e3 \\ \hline
\textbf{Break step} & 4.5e4 & 6e4 & 4.5e4 \\ \hline
\end{tabular}
\end{table}

\subsection{Backtest Performance}

\begin{figure*}
    \centering
    \includegraphics[width=1\linewidth]{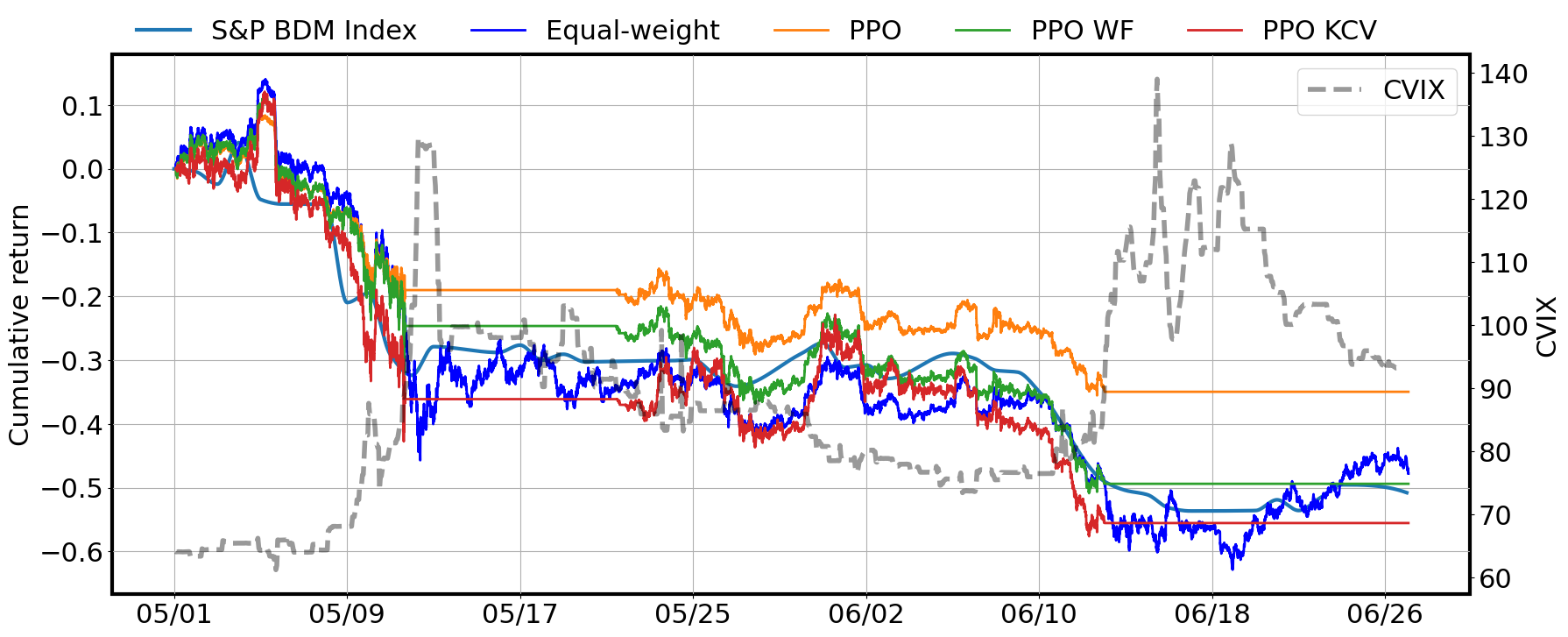}
    \caption{The average trading cumulative return curves for conventional agents. From 05/01/2022 to 06/27/2022, the initial capital is $\$1,000,000$.}
    \label{fig:test_cumulative_return53}
\end{figure*}

\begin{figure*}
    \centering
    \includegraphics[width=1\linewidth]{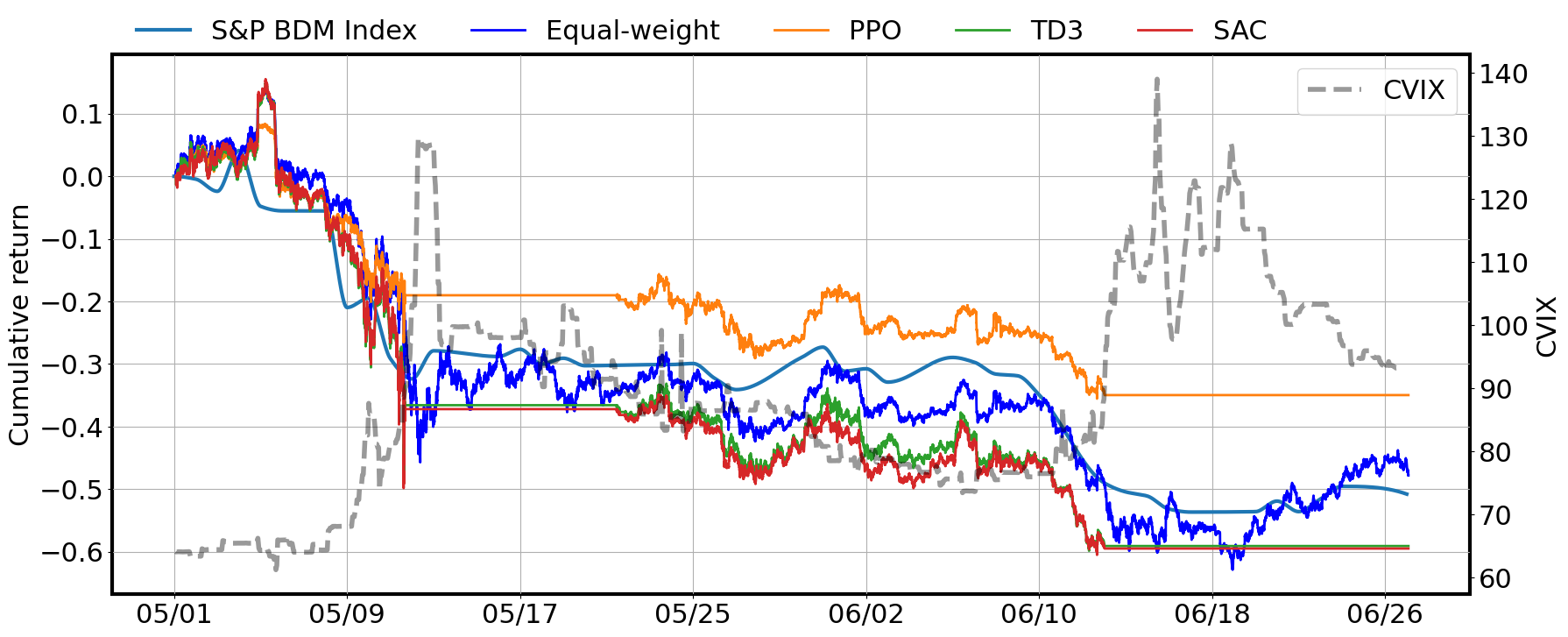}
    \caption{The average trading cumulative return curves for DRL algorithms. From 05/01/2022 to 06/27/2022, the initial capital is $\$1,000,000$.}
    \label{fig:test_cumulative_return54}
\end{figure*}

Fig. \ref{fig:test_cumulative_return53} and Fig. \ref{fig:test_cumulative_return54} show the backtest results. When the CVIX indicator surpasses $90$, the agent stops buying and sells all cryptocurrency holdings. Fig. \ref{fig:test_cumulative_return53} and Table \ref{tab:Performance evaluation comparison} compare conventional agents, market benchmarks and our approach. Compared to PPO WF and PPO KCV, our method outperforms the other two agents with at least a 15\% increase in the cumulative return. The lower volatility of PPO indicates that our method is more robust to risk. Fig. \ref{fig:test_cumulative_return54} and Table \ref{tab:Performance evaluation comparison} show the backtest results of the DRL agents. The cumulative return of the PPO agent is significantly better ($>24\%$) than those of agents TD3 and SAC. Also, in terms of volatility, the PPO agent is superior.

Finally, compared to the benchmarks, the performance of our approach is more excellent in terms of cumulative return and volatility. From our method and experiments, we conclude that the superior agent is PPO.

\begin{table*}
\centering
\caption{Performance comparison.}
\label{tab:Performance evaluation comparison}
\begin{tabular}{|l|l|l|l|l|l|l|l|}
\hline
\textbf{Metrics $\times$ Method} & \textbf{S\&P  BDM Index} & \textbf{Equal-weight} & \textbf{PPO WF} & \textbf{PPO  KCV} & \textbf{PPO} & \textbf{TD3} & \textbf{SAC} \\ \hline
\textbf{Cumulative return} & $-50.78\%$ & $-47.78\%$ & $-49.39\%$ & $-55.54\%$ & $-34.96\%$ & $-59.08\% $ & $-59.48\%$ \\ \cline{1-1}
\textbf{Volatility} & $5.81\mathrm{e}{-2}$ & $4.19\mathrm{e}{-3}$ & $2.79\mathrm{e}{-3}$ & $3.49\mathrm{e}{-3}$ & $2.01\mathrm{e}{-3}$ & $3.51\mathrm{e}{-3}$ & $3.78\mathrm{e}{-3}$ \\ \cline{1-1}
\textbf{Prob. of overfitting $p$} & \multicolumn{1}{c|}{-} & \multicolumn{1}{c|}{-} & $17.5\%$ & $7.9\%$ & $8.0\%$ & $9.6\%$ & $21.3\%$ \\ \hline
\end{tabular}
\end{table*}


\section{Conclusion}
\label{section: Conclusion}

In this paper, we have shown the importance of addressing the backtesting overfitting issue in cryptocurrency trading with deep reinforcement learning. Results show that the least overfitting agent PPO (with combinatorial CV method) outperforms the conventional agents (WF and KCV methods), two other DRL agents (TD3 and SAC), and the S\&P DBM Index in cumulative return and volatility, showing good robustness.

Future work will be interesting to 1). explore the evolution of the probability of overfitting during training and for different agents; 2). test limit order setting and trade closure; 3). explore large-scale data, such as all currencies corresponding to the S\&P BDM index; and 4). consider more features for the state space, including fundamentals and sentiment features.


\bibliography{aaai22}

\end{document}